\documentclass[prb,twocolumn,preprintnumbers,amsmath,amssymb,floatfix]{revtex4-1}

\usepackage{graphicx}
\usepackage{amssymb}
\usepackage{amsmath}
\usepackage{dcolumn}
\usepackage{bm}
\usepackage{color}

\makeatletter 
\def\@hangfrom@section#1#2#3{\@hangfrom{#1#2#3}}
\makeatother

\makeatletter        
\def\@biblabel#1{[#1]}
\makeatother

\newcommand{\cmt}[1]{} 
\newcommand{\hlt}[1]{#1} 

\usepackage[normalem]{ulem}
\usepackage{xcolor}
\renewcommand\sout{\bgroup\markoverwith{\textcolor{red}{\rule[0.5ex]{2pt}{0.4pt}}}\ULon}

\mathchardef\mhyphen="2D

\usepackage{hyperref}
\begin{document}


\title{Critical dehydrogenation steps of perhydro-$N$-ethylcarbazole on Ru(0001) surface}


\author{Chunguang Tang$^{1,2}$}
\email{chunguang.tang@anu.edu.au}
\author{Preetham Permude$^{1}$}
\author{Shunxin Fei$^3$}
\author{Terry J. Frankcombe$^4$}
\author{Sean C. Smith$^{2,5}$}
\author{Yun Liu$^{1,2}$}
\email{yun.liu@anu.edu.au}
\affiliation{$^1$ Research School of Chemistry, The Australian National University, Canberra, Australia;\\ $^2$ Institute of Climate, Energy and Disaster Solutions, The Australian National University, Canberra, Australia;\\ $^3$ School of Materials Science and Engineering, Anhui University of Technology, Maanshan 243002, People's Republic of China;\\$^4$ School of Physical, Environmental and
Mathematical Sciences, University of New South Wales,
Canberra, Australia.\\$^5$ Research School of Physics, The Australian National University, Canberra, Australia}

\begin{abstract}

Understanding of the critical atomistic steps during the dehydrogenation process of liquid organic hydrogen carriers (LOHCs) is important to the design of cost-efficient, high-performance LOHC catalysts. Based on the density functional theory (DFT) we studied the thermodynamics and kinetics of the complete dehydrogenation path of perhydro-$N$-ethylcarbazole (12H-NEC) on Ru(0001) surface, involving the adsorption of 12H-NEC, the discharge of H ions onto Ru surface, and the desorption of H$_2$ and hydrogen-lean NEC. It was found that the bonding of $n$H-NEC is significantly strengthened for $n$$\leq$4 because of the flat aromatic ring. Although the whole dehydrogenation process is endothermic, the release of H from $n$H-NEC, with H adsorbed onto the Ru surface, was found to be exothermic. The desorption of flat, hydrogen-lean NEC, which costs $\sim$255 kJ/mol, was identified as the most energy demanding step. In addition, the effect of surface morphology on adsorption was studied based on an amorphous surface model. Overall, the results imply more efficient dehydrogenation could be achieved from relatively weak bonding of NEC to catalysts, either through engineering catalyst surface (such as surface defects or smaller catalyst particles) or different catalyst materials. Our calculations also revealed possible dealkylation at elevated temperatures.

\vspace{10pt}
Published version at \href{https://doi.org/10.1016/j.commatsci.2023.112373}{Computational Materials Science, 229, 112373 (2023)}

\end{abstract}
\maketitle

\section{Introduction}

Liquid organic hydrogen carriers (LOHCs) have attracted extensive research interests \cite{sotoodeh_overview_2013, preuster_liquid_2017, modisha_prospect_2019} as a potential alternative hydrogen storage approach to the traditional approaches such as liquid H$_2$, compressed H$_2$ gas and even competitive circular carriers \cite{abdin_large-scale_2021}. Nevertheless, the large-scale application of LOHCs faces challenges such as abundant, cost-effective LOHCs as well as relevant catalysts. Currently a number of LOHCs, including benzene, toluene, naphthalene and $N$-ethylcarbazole (NEC), have been identified as potential LOHCs and their properties were discussed in some review or perspective articles  \cite{modisha_prospect_2019, niermann_liquid_2019, abdin_large-scale_2021}. Exploration of new LOHCs, such as those from natural products, has also been carried out recently \cite{tang_natural_2020}. For a given LOHC material, catalysts significantly impact on its (de)hydrogenation process and hence represent a key parameter for LOHC performance. To date, a number of catalysts have been explored for the above-mentioned LOHCs, as summarized in recent reviews \cite{modisha_prospect_2019}. For example, for dehydrogenation of cyclohexane, the performance of Ni-based \cite{biniwale_dehydrogenation_2005, xia_cyclohexane_2017}, Ag-based \cite{pande_catalytic_2012}, Pt-based \cite{kariya_efficient_2002, kariya_efficient_2003} catalysts with various supports has been investigated and the synergistic effect from the addition of a second metal has been reported.  

As an ideal tool for investigating atomistic process and complementary to experiments, first principles modelling based on the density functional theory (DFT) has been often used to provide fundamental insights into the LOHC-catalyst reactions \cite{shin_thermodynamic_2018, yook_density_2020, ouma_insight_2018}. For example, DFT calculations have been used to examine the structural features of tetrahydrocarbazole adsorption on Pd surfaces and its preferred dehydrogenation pathways \cite{crawford_understanding_2007}. The calculations of dodecahydro(12H)-NEC, 12H-carbazole, and 12H-fluorene on Pd(111) surface \cite{sotoodeh_dehydrogenation_2012} have revealed a linkage between the dehydrogenation rate and the adsorption strength. Some researchers have explored the adsorption of NEC and its hydrogenated states on different Ru surfaces \cite{eblagon_study_2010} and suggested that intermediate 8H-NEC, which is kinetically stable, prefers surface sites of low coordination (such as edges). To date most studies have focused on the adsorption of LOHCs on catalyst surface, but the atomistic-scale (de)hydrogenation process involves multiple subprocesses, such as (for dehydrogenation) the adsorption of hydrogen-rich LOHCs on catalyst surface, the release of H onto catalyst surface, the recombination of H ions into H$_2$ molecules and the desorption of H$_2$ molecules and hydrogen-lean LOHCs. Therefore, understanding of the full picture of the dehydrogenation process is necessary for identifying the critical steps, which is important for designing cost-efficient, high-performance catalysts for LOHCs. In this work we compare the adsorption/desorption energetics as well as kinetics of both NEC-based molecules and hydrogen species on Ru(0001) in an attempt to identify the energetically critical steps for dehydrogenating 12H-NEC. We also examine the impact of imperfect surfaces on the adsorption of the molecules based on an amorphous Ru structure. In addition, a critical challenge for controlling dehydrogenation is the potential fluctuation of temperature in a big reactor, which may cause the decomposition of LOHCs \cite{gleichweit_dehydrogenation_2013}, especially when the LOHCs contain weak bonds. In this work we also address the dealkylation of NEC and compare with the other events in energetics.

\begin{table*}
\caption{Calculated dehydrogenation energy (kJ/mol-H$_2$) based on PBE and hybrid HSE03 functionals, with and without van der Waals (VDW) and thermal corrections. Thermal corrections and experimental data are for the standard condition (1 atm and 298 K). More details are in work \cite{tang_natural_2020} and references therein. $\delta E=\Delta E_{\rm{PBE}}-\Delta E_{\rm{HSE03}}^{\rm{thermal}}$. We note $\Delta E_{\rm{PBE}}$ for perhydro-NEC in this table (63.9 kJ/mol-H$_2$) is slightly different from that (401.48/6=66.9 kJ/mol-H$_2$) in Table \ref{tab:de}. The difference may arise from different parameters used in the previous work \cite{tang_natural_2020} and this study, such as the perhydro-NEC isomer chosen and the planewave cutoff energy.}
\setlength{\tabcolsep}{5.2pt}
\begin{tabular} {lcccccccc}  \hline  \hline
\multicolumn{9}{c}{Dehydrogenation energy averaged over reaction steps}\\
Hydrogen-rich system & $\Delta H^{\circ}_{\rm{expt}}$ &$\Delta E_{\rm{PBE}}$ &$\Delta E_{\rm{HSE}}$ &$\Delta E_{\rm{HSE+VDW}}$ & $\Delta E_{\rm{PBE}}^{\rm{thermal}}$ & $\Delta E_{\rm{HSE}}^{\rm{thermal}}$ & $\Delta E_{\rm{HSE+VDW}}^{\rm{thermal}}$& $\delta E$ \\
\hline
Cyclohexane &68.7 & 79.36 & 92.29 & 96.26& 54.62 &67.56 &71.52&11.8 \\
Decalin &66.55& 84.08 & 97.88 &102.52& 57.13 &70.93 &75.57&13.2 \\
Perhydro-NEC  &53.2& 63.94 &77.82 &83.13 & 38.14&52.02 &57.33&11.9 \\
Methylcyclohexane  &68.3& 87.28 & 100.85 & 105.25& 61.25&74.82 &79.22 &12.5\\
Perhydro-trisphaeridine & &66.40 &81.11 & & 39.83 & 54.54 & &12.8\\
\hline\hline
\end{tabular}
\setlength{\tabcolsep}{17.75pt}
\begin{tabular}{lccccccc}
\multicolumn{8}{c}{Stepwise dehydrogenation energy for perhydro-trisphaeridine}\\
Energy & step 1 &  step 2 & step 3 & step 4 & step 5 & step 6 & step 7 \\
\hline
$\Delta E_{\rm{PBE}}$ & 60.94& 106.98 & -7.34 & 103.52 & 42.54 & 129.10 & 29.08 \\
$\Delta E_{\rm{HSE}}^{\rm{thermal}}$& 47.77 &  96.24 & -21.41 & 94.67 & 31.12 & 117.14 & 16.25 \\
$\delta E$ & 13.17 & 10.74 & 14.07 & 8.85 & 11.42 & 11.96 & 12.83\\
\hline\hline
\end{tabular}
\label{tab:natural}
\end{table*}
\vspace{20pt}

\begin{figure}
\includegraphics[width=3.5in]{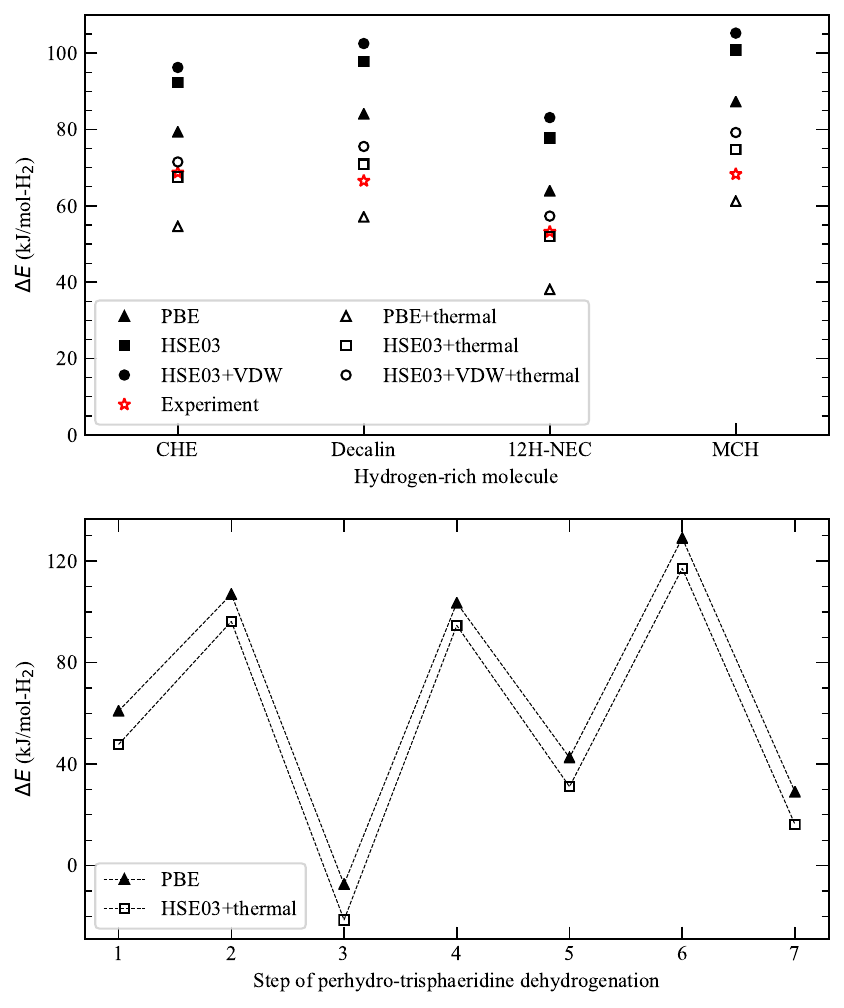}
\caption{Comparison of the accuracy of various  approximation methods used for dehydrogenation energy calculation. (top) Calculated dehydrogenation energy (kJ/mol-H$_2$) of various molecules with different levels of approximation, as compared with experimental values. The D3 method with Becke-Jonson damping was used for van der Waals (VDW) correction. (bottom) Stepwise dehydrogenation energy for perhydro-trisphaeridine. Data from Table \ref{tab:natural}. }
\label{fig:natural}
\end{figure}

\section{Methods}

Geometry optimization of all the relevant structures in this work was performed using the code VASP \cite{kresse_efficient_1996} with the GGA-PBE exchange-correlation functional and a plane wave basis set with an energy cutoff of 300 eV. For molecule-only structures, a simulation cell separating the molecules from their periodic images by more than 10 $ $\AA$ $ was used, and the $k$-point sampling was performed at the $\Gamma$ point. For bulk Ru, an orthogonal cell ($\sim$2.7$\times$4.69$\times$4.28 $ $\AA$^3$) with 11$\times$7$\times$7 $k$-point sampling was used for structural optimization. To represent the catalyst surface, we built an orthogonal Ru(0001) (7$\times$8) slab out of the optimized bulk Ru structure with 3 atomic layers in thickness, in view of the balance of accuracy and computational cost. The molecule/slab structure was then contained in a simulation cell of $\sim$18.8$\times$18.9$\times$25.5 $ $\AA$^3$, which separates the structure with its periodic images along the surface normal by $\sim$15 $ $\AA$ $ and is coupled with $k$-point sampling at the $\Gamma$ point. During the calculations the bottom Ru layer was fixed. The geometry optimization of the system was stopped after the force on atoms is below 0.01 eV/$ $\AA$ $.

To examine the effect of an imperfect Ru surface on adsorption, we first constructed an amorphous bulk Ru based on classical molecular dynamics, with the atomic interactions described by an embedded atom method (EAM) potential \cite{fortini_asperity_2008}. To this end, a bulk Ru of 224 atoms (similar to the slab structure mentioned above but with 4 layers) was well liquefied at 2500 K and then quenched to 100 K at 50 K/ps. The simulations were carried out under an NPT (constant particle number, pressure, and temperature) ensemble using code LAMMPS \cite{plimpton_fast_1995} with Nos$\acute{\rm{e}}$-Hoover thermostat and barostat, and the timestep was set as 2 fs. The obtained amorphous structure, after being optimized using DFT, was used to create a slab by inserting a vacuum layer of $\sim$15 $ $\AA$ $ into the simulation cell, with the final cell size being $\sim$19.4$\times$18.7$\times$25.5 $ $\AA$^3$. The $n$H-NEC molecules were then put to 9 different surface sites of the slab for geometry optimization, with the bottom Ru layer of thickness $\sim$3.5  $ $\AA$ $ being fixed.

A previous study on natural LOHCs \cite{tang_natural_2020} indicates that DFT calculations using a PBE-based hybrid functional (HSE03) plus thermal corrections, which include the zero point energy and the enthalpy of the molecules, can predict the dehydrogenation enthalpy reasonably well. However, the thermal corrections result in extra computational cost and, more importantly, sometimes we experienced difficulties for the hybrid functional calculations to converge, especially for metals. To examine the feasibility of studying the dehydrogenation reactions using the PBE functional alone, we tested the performace of PBE and HSE03, with or without thermal and van der Waals corrections, on several example molecules following the parameters in reference \cite{tang_natural_2020}. As shown in Table \ref{tab:natural} and Fig. \ref{fig:natural}, although the PBE functional alone overestimates the average and stepwise dehydrogenation enthalpies by $>$10 and $\sim$9$-$14 kJ/mol-H$_2$, respectively, it provides very good estimates for the relative energetics. Hence, in this work we address the dehydrogenation energetics only based on the PBE functional at zero temperature.

\begin{table*}[t]
\caption{System energy (kJ/mol) calculated with the PBE functional at each step ($n$) of the four reactions shown in Fig. \ref{fig:de}. For reaction 1, we also computed the energy with the wB97XD functional and cc-pVTZ basis set for comparison, implemented with code Gaussian. $E_d$ represents the desorption energy of $n$H-NEC.}
\setlength{\tabcolsep}{15.5pt}
\begin{tabular} {llllllll}  \hline  \hline
reaction & $n=12$ & $n=10$  & $n=8$ &  $n=6$  &   $n=4$   &   $n=2$  &  $n=0$ \\
\hline
1 &0 & 97.55 & 120.22 & 248.55 &258.59 &392.61 &401.48 \\
1 (wB97XD) &0 & 122.18& 157.39& 309.08& 332.70& 488.99& 509.77\\
2&-66.45& 51.07& 76.55& 197.93& 91.11& 225.52& 146.50\\
3&-66.45& -5.76& -37.12& 27.43& -136.21& -58.63& -254.98\\
4&-66.45& -71.56& -168.92& -170.27& -399.81& -388.14& -589.89\\
\hline
$E_d^{n\rm{H-NEC}}$ & 66.45 & 46.48 & 43.67&50.62&167.48&167.09&254.98\\
\hline\hline
\end{tabular}
\label{tab:de}
\end{table*}

\begin{figure*}[htb]
\includegraphics[width=6in]{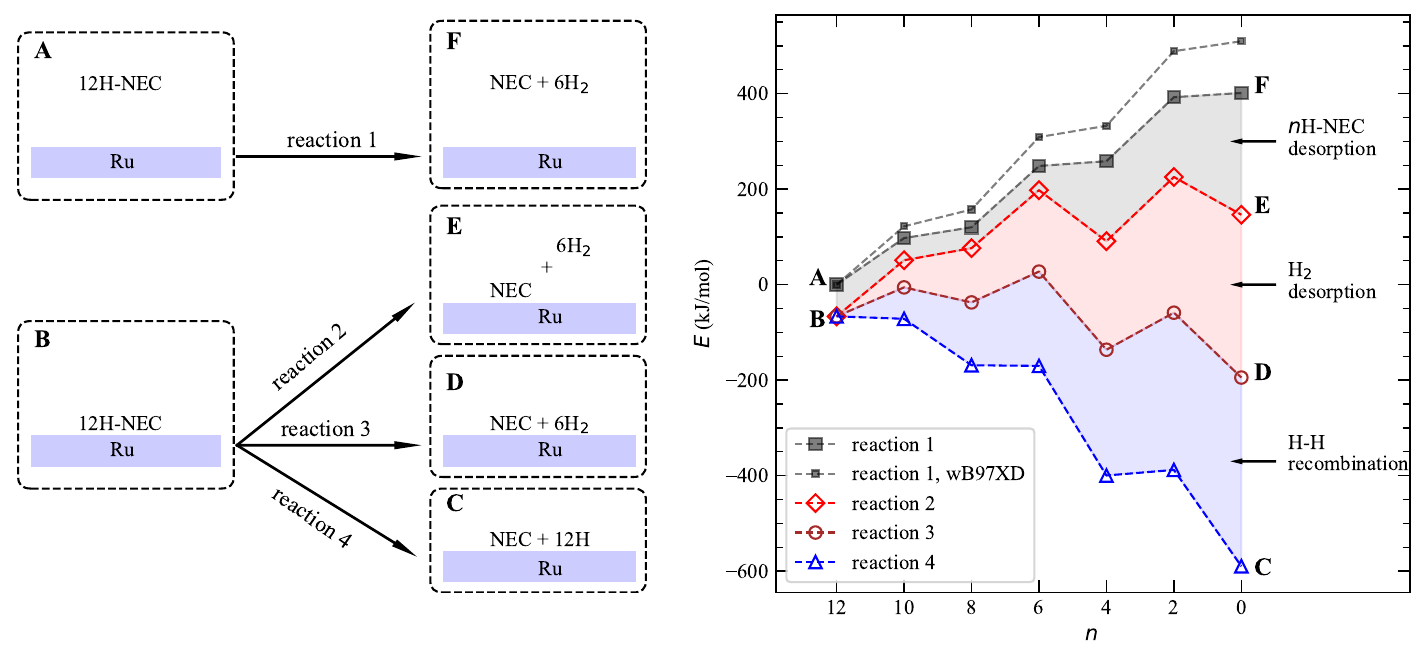}
\includegraphics[width=6in]{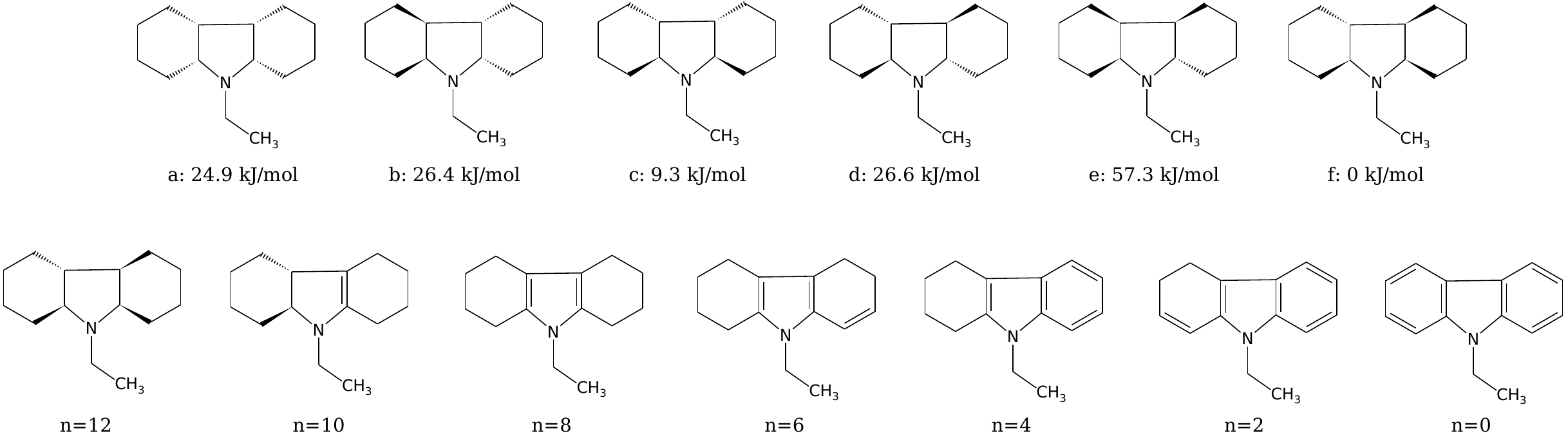}
\caption{Dehydrogenation of 12H-NEC on Ru(0001) surface. (top left) Schematic illustration of four reaction paths related to the dehydrogenation. (top right) System energy along the four reactions as a function of $n$, the number of H atoms charged to NEC, with the energies referenced to state A. For reaction 1, we also computed the energy with the wB97XD functional using code Gaussian for comparison. The energy values are in Table \ref{tab:de}. (bottom) Structures of 12-NEC isomers (a-f) with energy referenced to the most stable isomer and the structures of $n$H-NEC considered during dehydrogenation.}
\label{fig:de}
\end{figure*}

For the dehydrogenation process, we assume the initial state (denoted as state A) as a 12H-NEC molecule far from the Ru surface and the final state (state F) as a NEC molecule plus 6 H$_2$ molecules far from the Ru surface, as schematically shown in Fig. \ref{fig:de}. For energetics study, we consider the following intermediate states: 12H-NEC being adsorbed onto the Ru surface (state B), all of the 12 H ions being released (state C), and the H ions combining into 6 H$_2$ molecules on the surface (state D), the H$_2$ molecules being desorbed (state E), and the system reaches state F upon NEC desorption. Alternative to the process C\textrightarrow D\textrightarrow E, direct desorption of H ions into gaseous H$_2$ (C\textrightarrow E) is also possible, of which the energetics is straightforward. We note that the above order of states is assumed for convenience (for example, experimentally H$_2$ desorption may start while some remaining H ions are not released yet, or NEC desorption may occur before H$_2$ desorption is finished) and it does not affect the conclusions of this work. To identify the effects of various reaction processes and the catalyst on the dehydrogenation energetics, we computed the system energies along four reaction paths as detailed below.

\begin{figure*}[htb]
\includegraphics[width=3.5in]{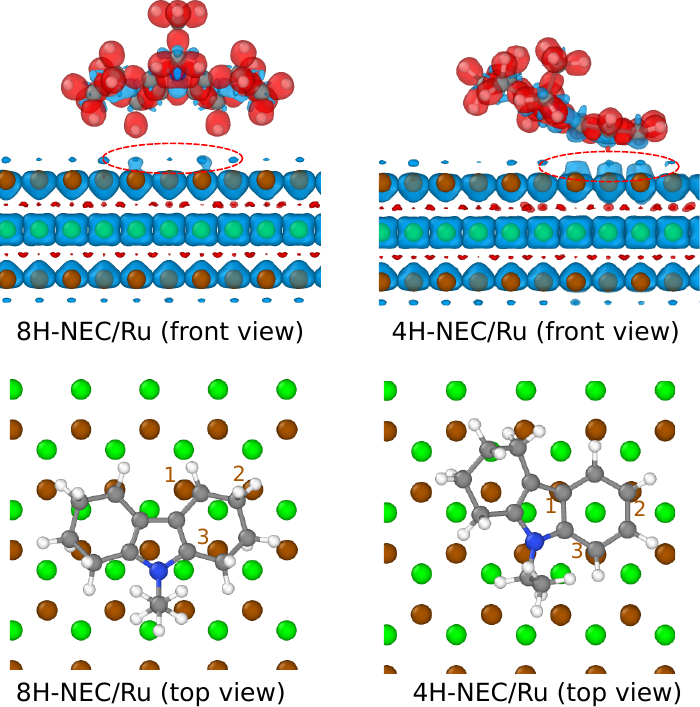}
\includegraphics[width=3.5in]{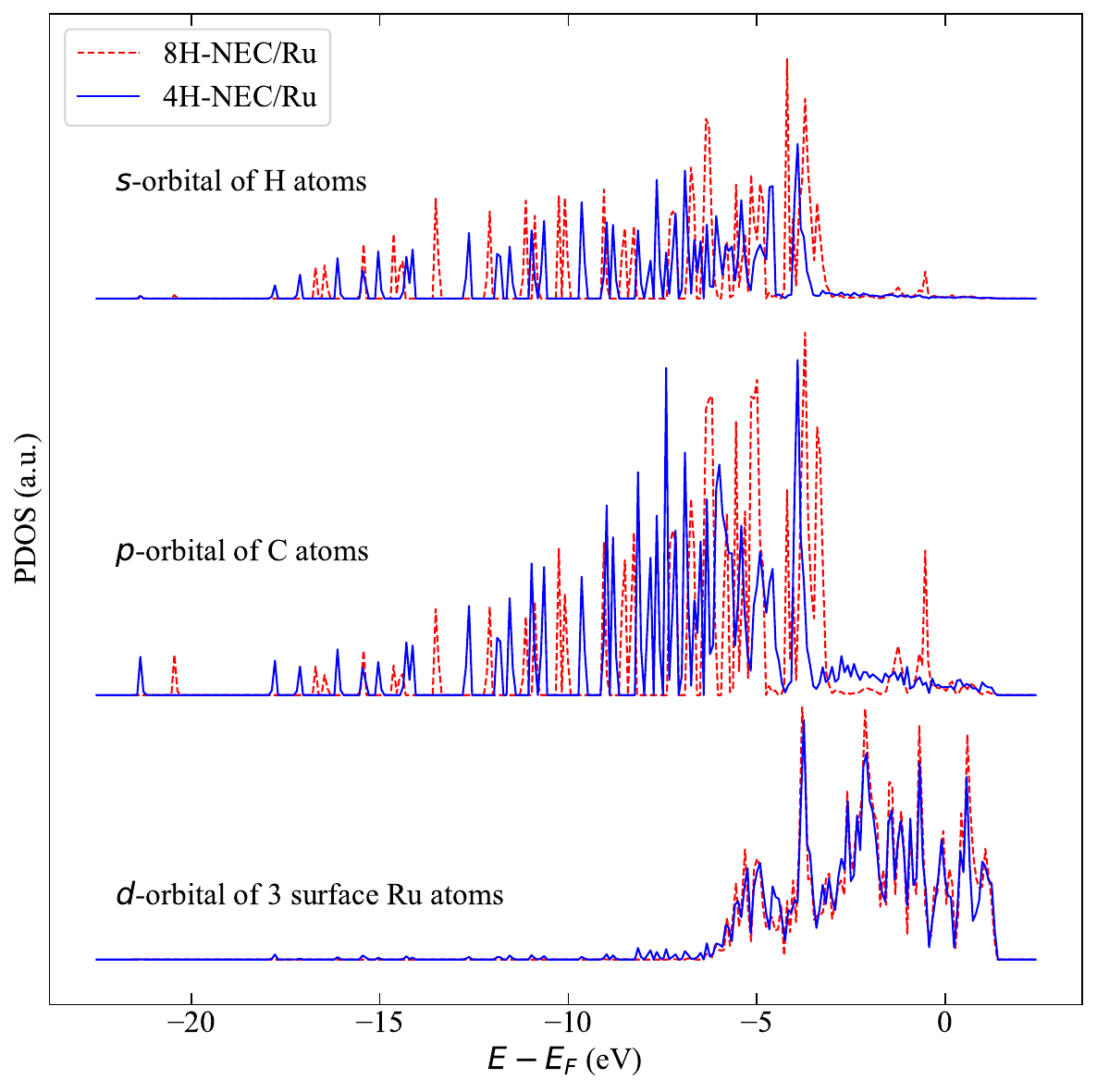}
\caption{Electronic structures of 8H-NEC and 4H-NEC on Ru(0001). (left) Structure and isosurface of the differential charge density, or the charge density minus the superposition of atomic charge densities. The isosurfaces represent charge density level of 10 $\mu e /$\AA$^3$ (red) and $-10~\mu e /$\AA$^3$ (blue). The brown and green spheres represent different Ru layers, and C, N, and H atoms are shown in grey, white, and blue, respectively. (right) Projected density of states of H and C atoms as well as three surface Ru atoms (labelled as 1-3 in the left).}
\label{fig:charge}
\end{figure*}

Reaction 1 (state A to state F) was considered for a reference scenario where Ru is not involved in the dehydrogenation process. In this case we treated the molecules and Ru slab separately and at each reaction step the system energy is the sum of the separate parts, which can be written as
\begin{equation}
E(n)=E_{n{\rm{H\mhyphen NEC}}} + \frac{12-n}{2}E_{\rm{H}_2} + E_{\rm{Ru}}
\end{equation}
where $n$ decreases from 12 for the initial state A to 0 for the final state F.

For reaction 2, we considered the transition from state B to state E, with  the released H ions directly forming gas H$_2$ molecules far from Ru surface. This hypothetical case is similar to reaction 1 except the $n$H-NEC molecule is adsorbed on Ru surface, and so it allows for identifying the effect of $n$H-NEC adsorption alone on dehydrogenation energetics. In this case, the system energy is 
\begin{equation}
    E(n)=E_{n{\rm{H\mhyphen NEC/Ru}}} + \frac{12-n}{2}E_{\rm{H}_2}
\end{equation}

Similarly, we considered transitions from B to D (reaction 3) and from B to C (reaction 4), with the released H ions forming adsorbed H$_2$ molecules and adsorbed H ions, respectively. This allows for identifying the adsorption effects of H$_2$ and H ions. Here we treat H$_2$ (or H ions) and the $n$H-NEC separately, and hence for reaction 3, the system energy can be written as 
\begin{equation}
 E(n)=E_{n{\rm{H\mhyphen NEC/Ru}}} + \frac{12-n}{2}\mu_{_{\rm{H_2}}}
\end{equation}
where the chemical potential $\mu_{_{\rm{H_2}}}$=$(E_{\rm{H}_2/Ru}$$-$$E_{\rm{Ru}})$ represents the energy of H$_2$ adsorbed on Ru surface. $E(n)$ for reaction 4 can be similarly defined.

The structures for $n$H-NEC we considered at various steps are shown at the bottom of Fig. \ref{fig:de}, which are suggested by a few previous studies \cite{fei_study_2017, dong_dehydrogenation_2016, yu_bimetallic_2020}. For 12H-NEC, there are six possible isomers\cite{eblagon_study_2010}, labelled as a-f in Fig. \ref{fig:de}, and we chose the one with the lowest calculated energy at 0 K. The chosen isomer is also one of the experimentally observed species \cite{eblagon_study_2010}.

\begin{figure}
\includegraphics[width=3.2in]{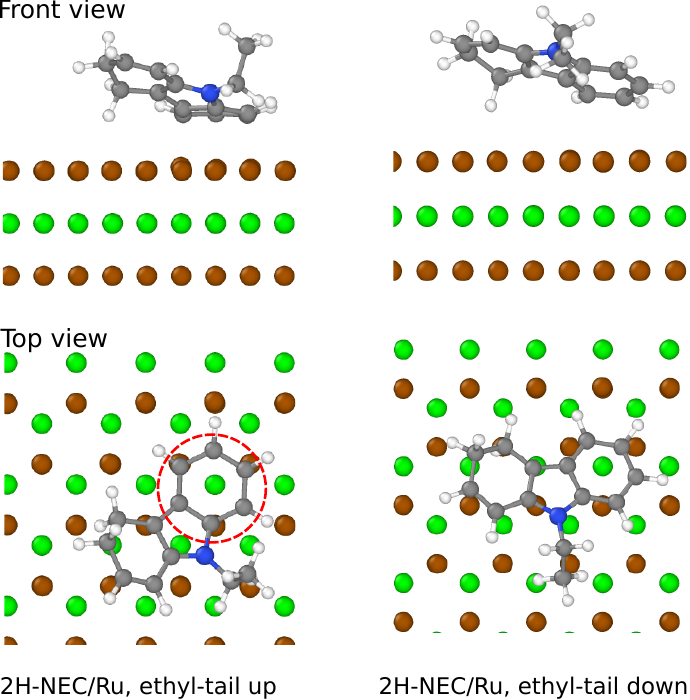}
\caption{Effect of ethyl group orientation on the adsorption of 2H-NEC on Ru(0001). The preferred ethyl-up state enables a better match of the aromatic ring to surface Ru.}
\label{fig:tail}
\end{figure}

\section{Results and discussions}
\subsection{Dehydrogenation thermodynamics}
The computed system energies for the four reactions, all referenced to that of state A, are listed in Table \ref{tab:de} and shown in Fig. \ref{fig:de}(right). For reaction 1, the calculated system energy increases as dehydrogenation proceeds, with the average energy increase to be $\sim$67 kJ/mol-H$_2$. For reaction 2, the system energies are reduced because of the adsorption of $n$H-NEC. The desorption energy ($E_d$) for $n$H-NEC, which is the $E$ difference between reactions 1 and 2, ranges from $\sim$44 to $\sim$66 kJ/mol for $n$$\geq$6 and from $\sim$167 to $\sim$255 kJ/mol for $n$$\leq$4. We attribute the significantly higher $E_d$ for $n$$\leq$4 to the formation of the aromatic ring, of which the flat structure allows for stronger C-Ru bonding with the flat Ru(0001) surface. To illustrate this we compare the the structures of 8H-NEC and 4H-NEC as an example. The former molecule adsorbs to Ru mainly via H atoms and hence the carbon rings are not as well aligned with the surface Ru atoms as the aromatic carbon ring in 4H-NEC (top view in Fig. \ref{fig:charge}). \hlt{The charge density isosurfaces in Fig. \ref{fig:charge} show} that more electrons transfer from surface Ru towards 4H-NEC than towards 8H-NEC because of the Ru-C bonding. \hlt{To quantitatively illustrate this, we performed Bader charge analysis. As shown in Table \ref{tab:bader}, the three Ru atoms beneath the aromatic ring of 4H-NEC lose $\sim$0.6 more electrons as compared to their counterparts beneath 8H-NEC.} Fig. \ref{fig:charge} also compares their projected density of states (PDOS) for H and C atoms as well as the three surface Ru atoms beneath the aromatic ring of 4H-NEC and the corresponding ring of 8H-NEC. By integrating over the PDOS, one can compute the electron energy of the corresponding orbitals (per atom) according to
\begin{equation}
E_e=\frac{1}{N_a} \int^{E_F}_{-\infty}n(\varepsilon)\varepsilon d\varepsilon
\end{equation}
where $n(\varepsilon)$ is the PDOS density, $E_F$ is the Fermi energy, and $N_a$ is the number of H, C, or Ru atoms contributing to the PDOS. We found $E_e$ values of $s_{_{\rm{H}}}$, $p_{_{\rm{C}}}$ and $d_{\rm{Ru}}$ of 4H-NEC/Ru are lower than those of 8H-NEC/Ru by $\sim$0.3, $\sim$1.2, and $\sim$1.3 eV, respectively. This indicates the stronger bonding between 4H-NEC and Ru, consistent with the charge density analysis.

\begin{table}[t]
\caption{\hlt{Bader charges of selected surface Ru atoms (refer to Fig. \ref{fig:charge}) and average Bader charges of the top surface Ru atoms. The charges, in unit $e$, are referenced to neutral Ru atoms. The analysis show that the top Ru layer gain electrons from the beneath Ru layer, resulting in a positive average charge.}}
\setlength{\tabcolsep}{6.5pt}
\begin{tabular} {ccccc}  \hline  \hline
System & atom 1 & atom 2  & atom 3 &  Surface Ru \\
\hline
8H-NEC/Ru &+0.11 & 0.00 & +0.04 & +0.07 \\
4H-NEC/Ru &$-$0.11 & $-$0.18& $-$0.15 & +0.06\\
\hline\hline
\end{tabular}
\label{tab:bader}
\end{table}

\begin{figure*}
\includegraphics[width=6.5in]{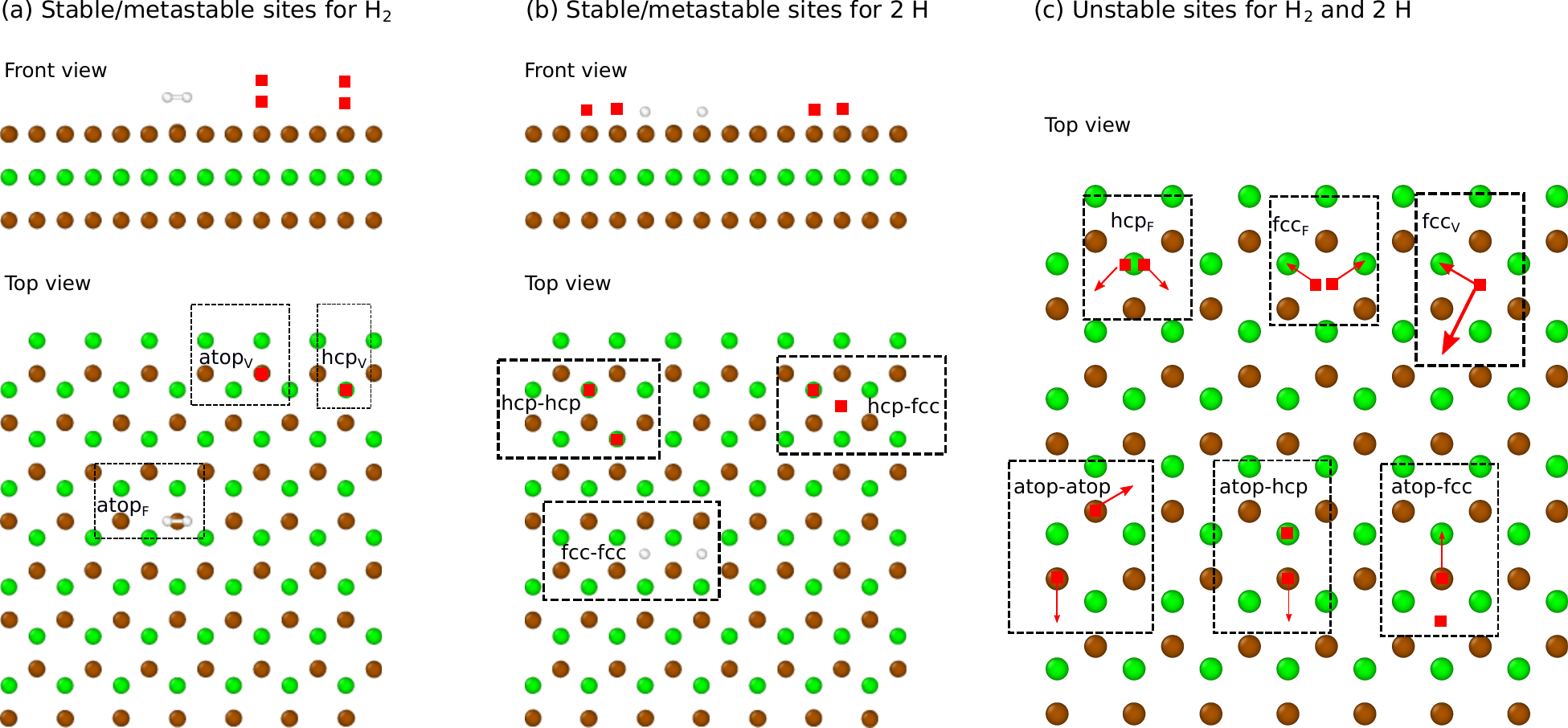}
\caption{Adsorption of H$_2$ molecule and 2 H atoms on Ru(0001) surface. In (a) and (b), the red squares schematically represent metastable adsorption sites. In (a), the adsorption energies ($E_{\rm{H_2/Ru}} -E_{\rm{Ru}}-E_{\rm{H_2}}$) were calculated to be $-$56.82, $-$2.47, and +0.78 kJ/mol for atop$_{_{\rm{F}}}$, atop$_{_{\rm{V}}}$, and hcp$_{_{\rm{V}}}$ site, respectively. The subscript F or V indicates a flat or vertical position of the H$_2$.  In (b), the adsorption energies ($E_{\rm{2H/Ru}} -E_{\rm{Ru}}-2$$\times$$E_{\rm{H}}$) are $-$548.73, $-$528.06, and $-$557.05 kJ/mol for hcp-hcp, hcp-fcc, and fcc-fcc site, respectively, which are close to previous calculations \cite{chou_theoretical_1989}. The adsorption energies become $-$114.36, $-$93.69, and $-$122.69 kJ/mol, respectively, if a H$_2$ molecule is assumed in the initial state (i.e., 2$\times$$E_{\rm{H}}$ is replaced by $E_{\rm{H_2}}$ in the formula). In (c) for the unstable sites, the squares/arrows represent the initial/final positions of hydrogen.}
\label{fig:h2ru}
\end{figure*}

For $n$H-NEC adsorption we tested the effect of the ethyl group state (pointing down to the surface or pointing up) by rotating the ethyl group around the neighboring N-C bond. We found that for $n\leq4$ the up state is highly favorable (by energy difference more than 70 kJ/mol) while for the other $n$H-NEC structures the two states are nearly equally favorable (with energy difference less than 10 kJ/mol). For 4H- and 2H-NEC, pointing-down of the ethyl group reduces the bonding of the molecule with Ru by displacing the aromatic ring from the favorable position (Fig. \ref{fig:tail}). \hlt{On the other hand, for $n$=0, the pointing-down ethyl group relaxes into a flat position, allowing the aromatic rings to stay parallel to the Ru surface, but we found that a pointing-up position is energetically more favorable.} For $n$$\geq$6, the charged H atoms increase the space between the carbon rings and the Ru surface, making the up/down state of the ethyl group relatively insignificant. \hlt{For gaseous $n$H-NEC molecules we did not find significant energy differences between the ethyl group states.} We note that our results are different from a previous study \cite{eblagon_study_2010} which reported negligible adsorption energy for 8H-NEC with the ethyl group facing down and $\sim$30 kJ/mol for facing up. However, both studies predict that the adsorption of 8H-NEC on a flat Ru surface is relatively weak, which could result in the desorption of 8H-NEC molecule. It was found that 8H-NEC prefers low-coordination sites, such as surface steps, to a flat surface \cite{eblagon_study_2010}.

\begin{figure*}
\includegraphics[width=2in]{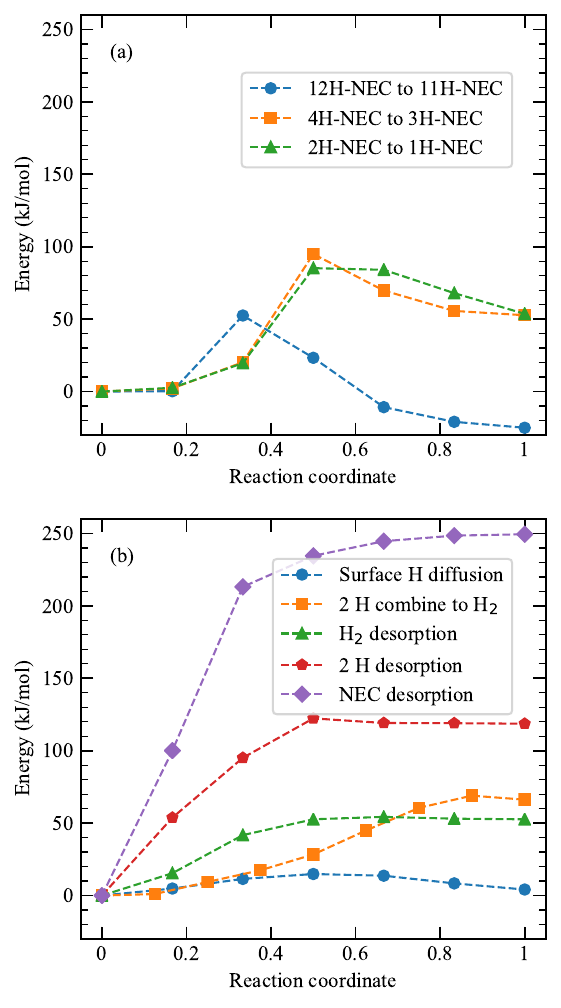}
\includegraphics[width=4.8in]{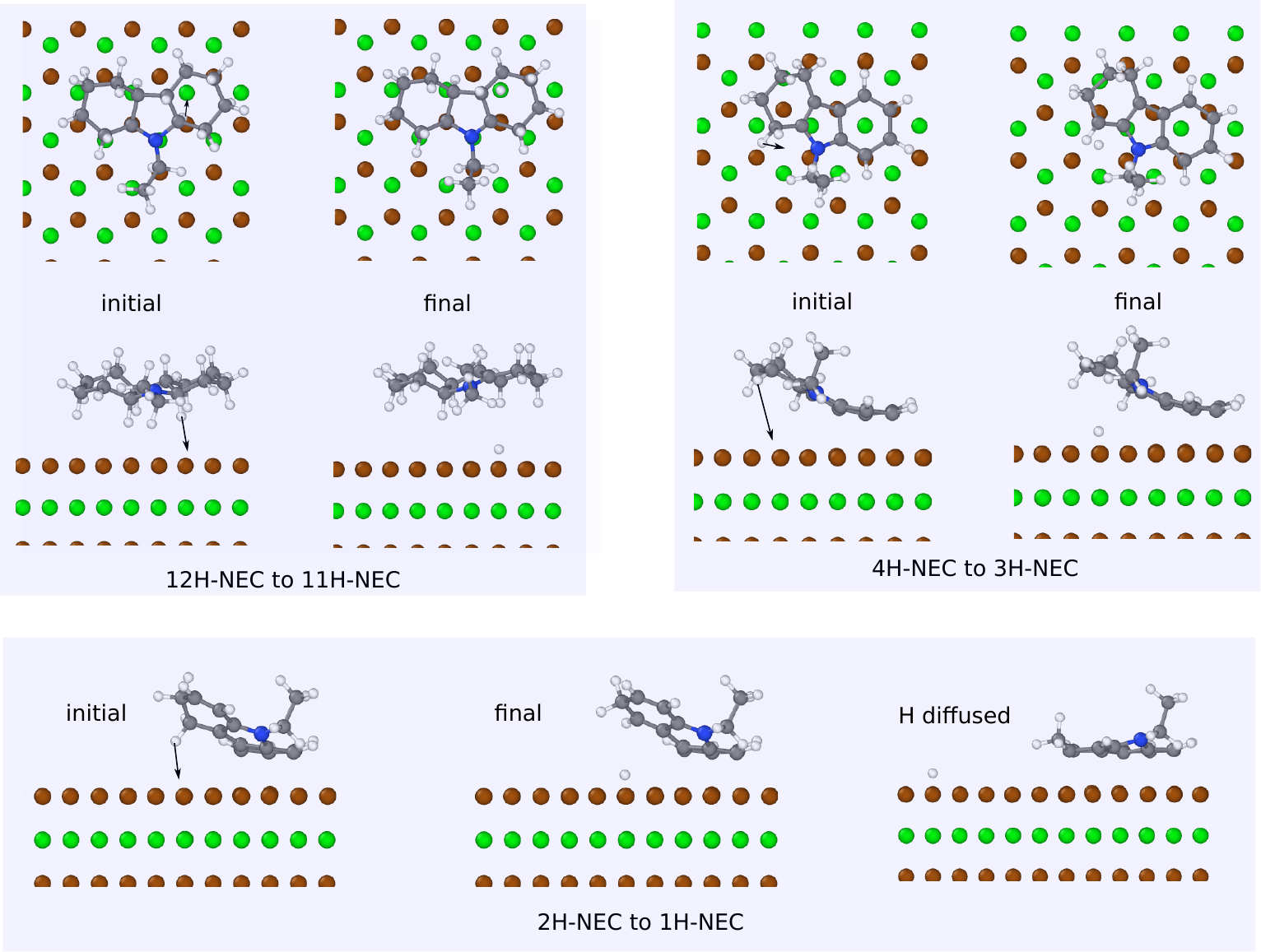}
\caption{Activation energies for dehydrogenation. (a) Hydrogen discharge from $n$H-NEC, with the reaction coordinate indicating the initial (0), intermediate, and final (1) states, and for each discharge event, the system energy is referenced to the initial state. The structure pictures are for the initial/final state of each discharge event. For 2H-NEC to 1H-NEC, the picture also shows further structural relaxation after the discharged H diffuses away. (b) Activation energies for other relevant surface events.}
\label{fig:neb}
\end{figure*}

If the released hydrogen are adsorbed onto the Ru surface, their chemical potential is lower than that of gas H$_2$ molecule and hence reduces the system energy. We studied the adsorption of hydrogen molecules and ions on various Ru surface sites (Fig. \ref{fig:h2ru}). It was found that hydrogen molecules are stable on the atop sites while two hydrogen ions could stay on hcp sites, fcc sites, or their combination, with the fcc sites being favorable. Assuming hydrogen chemical potential based on the most favorable hydrogen molecule/ion adsorptions, we obtained the system energies for reactions 3 and 4, respectively, as shown in Fig. \ref{fig:de}. These two reactions reveal an exothermic hydrogen release process, with the path of hydrogen ions being more favorable. 

Fig. \ref{fig:de} reveals not only the most probable path for 12-NEC dehydrogenation but also the contributions of different factors ($n$H-NEC desorption, H$_2$ desorption, and H ion recombination) to dehydrogenation energetics, which allows for identifying the energetically critical step. For example, for the dehydrogenation path A$\rightarrow$B$\rightarrow$C$\rightarrow$D$\rightarrow$E$\rightarrow$F, the first two steps are exothermic and the final three steps cost 335, 401, and 255 kJ/mol, respectively. In view that C$\rightarrow$D and D$\rightarrow$E are for 6 H$_2$ molecules while E$\rightarrow$F is the desorption of a single NEC molecule, it is clear that NEC desorption is the energetically critical step for the dehydrogenation. Because of the reversibility of (de)hydrogenation, it is easy to see that, for NEC hydrogenation (assuming the same catalyst), the endothermic step C$\rightarrow$B costs 523 kJ/mol, or on average $\sim$44 kJ/mol for charging each H atom, and the critical desorption of 12H-NEC costs $\sim$66 kJ/mol. As the lowest energy point in Fig. \ref{fig:de}, state C is also critical since too strong adsorption of H ions on the catalyst surface could be an energy trap that slows the reactions down. From Fig. \ref{fig:de} one also can predict the kinetically stable intermediate products, 8H-NEC and 4H-NEC, as observed in experiments \cite{dong_dehydrogenation_2016, jiang_experimental_2019}. Along path B$\rightarrow$C, the system energies of $n$=8 and $n$=4 are close to those of $n$=6 and $n$=2, respectively, which makes 8H-NEC and 4H-NEC kinetically stable in view of the activation energy for the reactions. On the other hand, the big thermodynamic driving force makes the other intermediate products ($n$=10, 6, 2) much less stable. Especially, we note that the formation of aromatic sextets (transitions $n$=6\textrightarrow$n$=4 and $n$=2\textrightarrow$n$=0) is associated with a big driving force ($\sim$200 kJ/mol). Similar phenomena were observed in our previous studies for bio-based LOHCs \cite{tang_natural_2020}.

Due to the strong NEC-Ru bonding and the relatively weak nitrogen-alkyl bonding, it is possible for dealkylation to occur before the desorption of NEC. Here we considered the dealkylation reaction by comparing the energy of NEC/Ru with two additional H atoms adsorbed on two fcc sites and that of a carbazole and an ethane adsorbed on Ru. It was found the latter is higher in energy by $\sim$\hlt{98.5} kJ/mol, much smaller than the desorption energy of NEC, implying the possibility of dealkylation under suitable situations. \hlt{Further, we note that the desorption of the resulting carbazole and ethane costs $\sim$212 kJ/mol, while the desorption energies of NEC and 2H (from two fcc sites to gaseous H$_2$) sum to $\sim$255+123=378 kJ/mol. This indicates that the dealkylation approach is thermodynamically favorable by 378$-$212$-$98.5$\approx$67 kJ/mol, which essentially is energy difference between (NEC+H$_2$) and (carbazole+ethane) in their gaseous state.} Experimentally, dealkylation of NEC on Pt(111) surface \cite{gleichweit_dehydrogenation_2013} was reported for temperatures above 390 K.  

\subsection{Dehydrogenation kinetics}
The above discussions reveal the thermodynamics during the (de)hydrogenation process of NEC on Ru. We also studied the kinetics of the process by computing the energy barrier of relevant steps on the Ru surface using the climbing image nudged elastic band method \cite{henkelman_climbing_2000}. The steps considered here include H discharge from $n$H-NEC, H diffusion, the combination of 2 H into a H$_2$ molecule, and the desorption of H$_2$ and NEC. For hydrogen discharge, although it is proper to consider H pairs in the above thermodynamics calculations and in theory it is possible for H to be discharged in pairs, here for the kinetics we consider a single H atom, consistent with experimental observations \cite{crawford_understanding_2007}. The H atoms on $n$H-NEC can be classified into three types, with the C-H bond roughly pointing down towards the Ru surface, parallel to the surface, or pointing up. We computed the activation energy for discharging a H atom from 12H-NEC, 4H-NEC, and 2H-NEC for example, with the corresponding C-H bond either pointing down or relatively parallel to the surface. As shown in Fig. \ref{fig:neb}(a), the activation energies for these cases range from $\sim$50 to $\sim$100 kJ/mol, which are similar to the calculated barriers ($\sim$39 to $\sim$92 kJ/mol) for dehydrogenating tetrahydrocarbazole to carbazole on palladium surface \cite{crawford_understanding_2007}. We note that for 4H-NEC$\rightarrow$3H-NEC and 2H-NEC$\rightarrow$1H-NEC, the final state is higher in energy than the initial state (Fig. \ref{fig:neb}(a)), partially because the relaxation of the system is limited when the discharged H atom sits below the $n$H-NEC. For 2H-NEC$\rightarrow$1H-NEC as an example, we found that once the discharged H atom diffuses away, the system decreases in energy by $\sim$130 kJ/mol with stronger bonding to the Ru surface.

For most 12H-NEC isomers (see Fig. \ref{fig:de}, bottom) adsorbed on the Ru surface, there exist some C-H bonds of the five-membered ring pointing up. The activation energy for discharging an upward H atom to Ru surface was found to be higher than 300 kJ/mol. Alternatively, we considered rotating the upward C-H bonds downwards, which allows easier H discharge later, and found similar activation energy. In view of the relatively low desorption energy of $n$H-NEC for $n$$\geq$6, a more probable mechanism for discharging these upward H atoms is that the $n$H-NEC molecule flips over the Ru surface and hence turns them downwards. In the presence of a solvent that accepts hydrogen, it is also possible to discharge the upward H atoms via the solvent \cite{crawford_understanding_2007}. 

We also computed the activation energies for other relevant events on Ru surface. For hydrogen migrating from a fcc site to a neighboring hcp site, the activation energies was found to be $\sim$14.9 kJ/mol only, which is consistent with the high mobility of H on Ru surface.  For the combination of two neighboring fcc-site H atoms into an atop H$_2$ molecule, $\sim$68.9 kJ/mol is necessary to activate the process. Finally, we examined the desorption of H$_2$ and NEC. For these events, we set the final states to be local energy minima corresponding to the H$_2$ and NEC molecules being $\sim$4 and $\sim$6 $ $\AA$ $, respectively, above the Ru surface. We note that these final states are slightly lower in energy than the states with the molecules infinitely away from the surface (by $\sim$4.2 and $\sim$5.5 kJ/mol for H$_2$ and NEC, respectively). Based on the selected final states, we computed the activation energies for H$_2$ and NEC desorption to be 54.3 and 249.5 kJ/mol, respectively, which are close to the desorption energies. 

In the above discussion of H$_2$ desorption, we assumed the path of two H atoms recombining into an atop H$_2$ before the desorption. The intermediate atop H$_2$ tends to decompose back into H atoms due to the small activation energy. Hence, we also studied the direct desorption of two H atoms from neighboring fcc sites into a gaseous H$_2$ molecule, as shown in Fig. \ref{fig:neb}(b). We found the activation energy to be $\sim$122 kJ/mol, close to the desorption energy. The calculated desorption energy in this work compares well to the experimental value, $\sim$120 kJ/mol, reported for low H coverage on Ru(0001) surface \cite{feulner_adsorption_1985}.

Overall, the kinetics calculations also identify the desorption of NEC as the critical step for dehydrogenation. On the other hand, the discharged H ions are found to be very mobile on Ru surface.
   
\subsection{Effect of surface morphology}
 
In the above we have discussed the dehydrogenation thermodynamics and kinetics of perhydro-$N$-ethylcarbazole on perfect Ru(0001) surface. In practice, a crystal surface may have structural defects and these defects may impact on the behavior of adsorbed molecules. For example, a previous study \cite{eblagon_study_2010} has shown that surface steps enables more stable adsorption of 8H-NEC as compared to a flat surface. Here, in a more general way, we use an amorphous Ru surface to represent an imperfect surface in view of its rich local structural features.  

\begin{figure}[h!]
\includegraphics[width=3.5in]{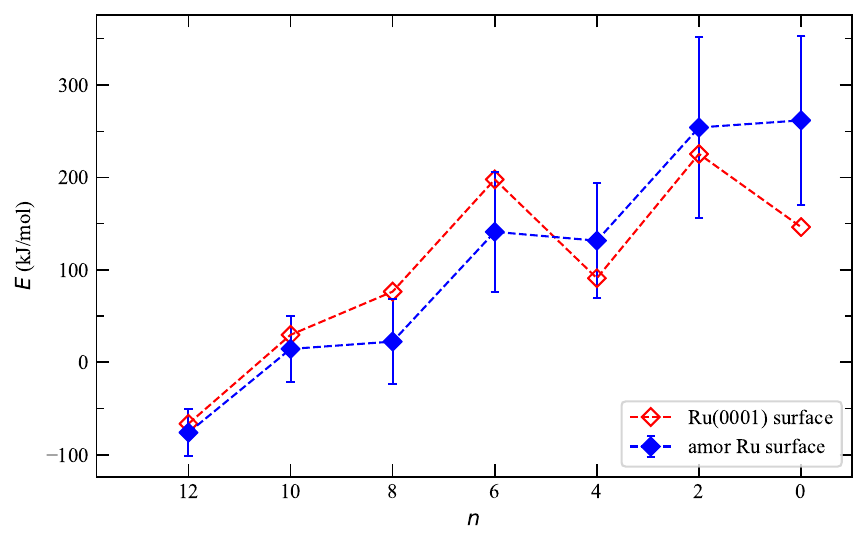}
\caption{System energy for reaction 2 on amorphous Ru surface and Ru(0001), with the energies referenced to state A for both cases. The energy for the amorphous case is averaged over nine different adsorption sites.}
\label{fig:amor}
\end{figure}

Fig. \ref{fig:amor} shows the average system energy for reaction 2 on the amorphous surface, as compared to that on Ru(0001). It can be seen that the adsorption energies of 12H- and 10H-NEC are similar for the two cases, but the energies for 8H- and 6H-NEC on the amorphous surface are about 50 kJ/mol lower than on Ru(0001). This highlights the impact of surface morphology and is qualitatively consistent with the stabilization effect of a surface step on 8H-NEC \cite{eblagon_study_2010}. The adsorption of 4H- and 2H-NEC may prefer a relatively flat surface due to the formation of the flat aromatic ring, and consequently the system energy for the amorphous case is slightly higher. This effect is more significant for the NEC case, with its adsorption energy on the amorphous surface is about 100 kJ/mol higher than on Ru(0001). Overall, however, the energy trends for reaction 2 in the two cases are similar, and the energy difference observed here does not alter the conclusions based on the crystalline phase.

\section{Summary}
Based on first principles calculations we studied the thermodynamics and kinetics of the complete dehydrogenation path of perhydro-$N$-ethylcarbazole (12H-NEC) on Ru(0001) surface, involving the adsorption of 12H-NEC, the release of H ions, H ion diffusion and recombination into H$_2$, and the desorption of H$_2$ and hydrogen-lean NEC. During the dehydrogenation process, the adsorption of $n$H-NEC on Ru(0001) is significantly strengthened upon the formation of aromatic ring, of which the flat structure allows for stronger C-Ru bonding. Although the whole dehydrogenation process is endothermic, the release of H from $n$H-NEC was found to be exothermic because of H adsorption onto Ru surface. The desorption of flat, hydrogen-lean NEC, which costs $\sim$255 kJ/mol, was identified as the most energy demanding step. Based on an amorphous model, we also showed that the surface morphology has a great impact on the stability of molecule adsorption. Overall, the results imply more efficient dehydrogenation could be achieved by weakening the bonding of NEC to catalysts, either through engineering catalyst surface (such as surface defects or smaller catalyst particles) or different catalyst materials. Our calculations also revealed possible dealkylation at elevated temperatures.

\vspace{10pt}
\noindent{\bf{Acknowledge}}
CT thanks the financial support from the Australian National University Grand Challenge program (Zero-Carbon Energy for the Asia-Pacific). This work was supported by computational resources provided by the Australian Government through the National Computational Infrastructure (NCI) under the ANU Merit Allocation Scheme.
 
\noindent{\bf{Data Availability}}
The raw/processed data required to reproduce these findings cannot be shared at this time as the data also forms part of an ongoing study.

\vspace{10pt}
 

\begin{thebibliography}{10}
\bibitem{sotoodeh_overview_2013} 
Sotoodeh F and Smith KJ, An overview of the kinetics and catalysis of hydrogen storage on organic liquids, {\textit{Can. J. Chem. Eng.}}, \textbf{91}, 1477, (2013).
\bibitem{preuster_liquid_2017} 
Preuster P, Papp C,  and Wasserscheid P, Liquid {Organic} {Hydrogen} {Carriers} ({LOHCs}): {Toward} a {Hydrogen} -free {Hydrogen} {Economy}, {\textit{Accounts Chem. Res.}}, \textbf{50}, 74, (2017).
\bibitem{modisha_prospect_2019} 
Modisha PM, Ouma CNM, Garidzirai R, Wasserscheid P,  and Bessarabov D, The {Prospect} of {Hydrogen} {Storage} {Using} {Liquid} {Organic} {Hydrogen} {Carriers}, {\textit{Energy Fuels}}, \textbf{33}, 2778, (2019).
\bibitem{abdin_large-scale_2021} 
Abdin Z, Tang C, Liu Y,  and Catchpole K, Large-scale stationary hydrogen storage via liquid organic hydrogen carriers, {\textit{iScience}}, \textbf{24}, 102966, (2021).
\bibitem{niermann_liquid_2019} 
Niermann M, Drünert S, Kaltschmitt M,  and Bonhoff K, Liquid organic hydrogen carriers ({LOHCs}) – techno-economic analysis of {LOHCs} in a defined process chain, {\textit{Energy Environ. Sci.}}, \textbf{12}, 290, (2019).
\bibitem{tang_natural_2020} 
Tang C, Fei S, Lin GD,  and Liu Y, Natural liquid organic hydrogen carrier with low dehydrogenation energy: {A} first principles study, {\textit{International Journal of Hydrogen Energy}}, \textbf{45}, 32089, (2020).
\bibitem{biniwale_dehydrogenation_2005} 
Biniwale RB, Kariya N,  and Ichikawa M, Dehydrogenation of cyclohexane over {Ni} based catalysts supported on activated carbon using spray-pulsed reactor and enhancement in activity by addition of a small amount of {Pt}, {\textit{Catal. Lett.}}, \textbf{105}, 83, (2005).
\bibitem{xia_cyclohexane_2017} 
Xia Z, Lu H, Liu H, Zhang Z,  and Chen Y, Cyclohexane dehydrogenation over {Ni}-{Cu}/{SiO2} catalyst: {Effect} of copper addition, {\textit{Catalysis Communications}}, \textbf{90}, 39, (2017).
\bibitem{pande_catalytic_2012} 
Pande JV, Shukla A,  and Biniwale RB, Catalytic dehydrogenation of cyclohexane over {Ag}-{M}/{ACC} catalysts for hydrogen supply, {\textit{International Journal of Hydrogen Energy}}, \textbf{37}, 6756, (2012).
\bibitem{kariya_efficient_2002} 
Kariya N, Fukuoka A,  and Ichikawa M, Efficient evolution of hydrogen from liquid cycloalkanes over {Pt}-containing catalysts supported on active carbons under "wet-dry multiphase conditions", {\textit{Appl. Catal. A-Gen.}}, \textbf{233}, 91, (2002).
\bibitem{kariya_efficient_2003} 
Kariya N $et~al$., Efficient hydrogen production using cyclohexane and decalin by pulse-spray mode reactor with {Pt} catalysts, {\textit{Appl. Catal. A-Gen.}}, \textbf{247}, 247, (2003).
\bibitem{shin_thermodynamic_2018} 
Shin BS, Yoon CW, Kwak SK,  and Kang JW, Thermodynamic assessment of carbazole-based organic polycyclic compounds for hydrogen storage applications via a computational approach, {\textit{Int. J. Hydrog. Energy}}, \textbf{43}, 12158, (2018).
\bibitem{yook_density_2020} 
Yook H, Kim K, Park JH, Suh Y,  and Han JW, Density functional theory study on the dehydrogenation of 1,2-dimethyl cyclohexane and 2-methyl piperidine on {Pd} and {Pt} catalysts, {\textit{Catal. Today}}, \textbf{352}, 345, (2020).
\bibitem{ouma_insight_2018} 
Ouma CNM, Modisha P,  and Bessarabov D, Insight into the adsorption of a liquid organic hydrogen carrier, perhydro-i-dibenzyltoluene (i = m, o, p), on {Pt}, {Pd} and {PtPd} planar surfaces, {\textit{RSC Adv.}}, \textbf{8}, 31895, (2018).
\bibitem{crawford_understanding_2007} 
Crawford P $et~al$., Understanding the dehydrogenation mechanism of tetrahydrocarbazole over palladium using a combined experimental and density functional theory approach, {\textit{J. Phys. Chem. C}}, \textbf{111}, 6434, (2007).
\bibitem{sotoodeh_dehydrogenation_2012} 
Sotoodeh F, Huber BJM,  and Smith KJ, Dehydrogenation kinetics and catalysis of organic heteroaromatics for hydrogen storage, {\textit{Int. J. Hydrog. Energy}}, \textbf{37}, 2715, (2012).
\bibitem{eblagon_study_2010} 
Eblagon KM $et~al$., Study of {Catalytic} {Sites} on {Ruthenium} {For} {Hydrogenation} of {N}-ethylcarbazole: {Implications} of {Hydrogen} {Storage} via {Reversible} {Catalytic} {Hydrogenation}, {\textit{J. Phys. Chem. C}}, \textbf{114}, 9720, (2010).
\bibitem{gleichweit_dehydrogenation_2013} 
Gleichweit C $et~al$., Dehydrogenation of {Dodecahydro}-{N}-ethylcarbazole on {Pt}(111), {\textit{ChemSusChem}}, \textbf{6}, 974, (2013).
\bibitem{kresse_efficient_1996} 
Kresse G and Furthmuller J, Efficient iterative schemes for ab initio total-energy calculations using a plane-wave basis set, {\textit{Phys. Rev. B}}, \textbf{54}, 11169, (1996).
\bibitem{fortini_asperity_2008} 
Fortini A, Mendelev MI, Buldyrev S,  and Srolovitz D, Asperity contacts at the nanoscale: {Comparison} of {Ru} and {Au}, {\textit{Journal of Applied Physics}}, \textbf{104}, 074320, (2008).
\bibitem{plimpton_fast_1995} 
Plimpton S, Fast {Parallel} {Algorithms} for {Short}-{Range} {Molecular}-{Dynamics}, {\textit{J. Comput. Phys.}}, \textbf{117}, 1, (1995).
\bibitem{fei_study_2017} 
Fei S $et~al$., A study on the catalytic hydrogenation of {N}-ethylcarbazole on the mesoporous {Pd}/{MoO3} catalyst, {\textit{Int. J. Hydrog. Energy}}, \textbf{42}, 25942, (2017).
\bibitem{dong_dehydrogenation_2016} 
Dong Y, Yang M, Mei P, Li C,  and Li L, Dehydrogenation kinetics study of perhydro-{N}-ethylcarbazole over a supported {Pd} catalyst for hydrogen storage application, {\textit{Int. J. Hydrog. Energy}}, \textbf{41}, 8498, (2016).
\bibitem{yu_bimetallic_2020} 
Yu H $et~al$., Bimetallic {Ru}-{Ni}/{TiO2} catalysts for hydrogenation of {N}-ethylcarbazole: {Role} of {TiO2} crystal structure, {\textit{Journal of Energy Chemistry}}, \textbf{40}, 188, (2020).
\bibitem{chou_theoretical_1989} 
Chou MY and Chelikowsky JR, Theoretical study of hydrogen adsorption on {Ru}(0001): {Possible} surface and subsurface occupation sites, {\textit{Phys. Rev. B}}, \textbf{39}, 5623, (1989).
\bibitem{jiang_experimental_2019} 
Jiang Z, Gong X, Wang B, Wu Z,  and Fang T, A experimental study on the dehydrogenation performance of dodecahydro-{N}-ethylcarbazole on {M}/{TiO2} catalysts, {\textit{International Journal of Hydrogen Energy}}, \textbf{44}, 2951, (2019).
\bibitem{henkelman_climbing_2000} 
Henkelman G, Uberuaga BP,  and Jonsson H, A climbing image nudged elastic band method for finding saddle points and minimum energy paths, {\textit{J. Chem. Phys.}}, \textbf{113}, 9901, (2000).
\bibitem{feulner_adsorption_1985} 
Feulner P and Menzel D, The adsorption of hydrogen on ruthenium (001): {Adsorption} states, dipole moments and kinetics of adsorption and desorption, {\textit{Surface Science}}, \textbf{154}, 465, (1985).
\end{thebibliography}
\end{document}